# The hyperfine properties of a hydrogenated Fe/V superlattice

M.Elzain, M. Al-Barwani, A.Gismelseed, A.Al-Rawas, A.Yousif, H.Widatallah, K.Bouziane and I.Al-Omari

*Department of Physics, College of Science, Box 36, Sultan Qaboos University, Al Khod 123, OMAN*

elzain@squ.edu.om

**Abstract**: We study the effect of hydrogen on the electronic, magnetic and hyperfine structures of an iron-vanadium superlattice consisting of three Fe monolayers and nine V monolayers. The contact charge density ($\rho$), the contact hyperfine field ($B_{hf}$) and the electronic field gradient (EFG) at the Fe sites for different H locations and H fillings are calculated using the first principle full-potential linear-augmented-plane-wave (FP-LAPW) method . It is found that sizeable changes in the hyperfine properties are obtained only when H is in the interface region.

## 1. Introduction

The magnetic nanostructures are currently part and parcel of the advanced technologies in magnetic applications [1]. In particular, magnetic superlattices are being used in magnetic recording based on their giant magneto-resistance attributes. The interlayer exchange coupling (IEC) of magnetic superlattices was found to be controllable by introduction of hydrogen in the superlattice systems [2]

Iron-vanadium superlattices constitute a benchmark for the various studies carried out on magnetic nanostructures. Due to the slight mismatch between the Fe and V lattice constants, the superlattices have tetragonal structures. The *c/a* ratio was found to depend on the thicknesses of Fe and V layers [3]. A giant magnetoresistance in Fe/V superlattices was reported by Broddefalk et. al. [4]. The quality of the Fe-V interface of dc sputtered samples was studied by Andersson et al [5] using conversion electron Mössbauer spectroscopy (CEMS) and x-ray diffraction. Sharp interfaces were found for growth temperatures of $330^o$ C or less. The distribution of the magnetic hyperfine field shows distinct peaks that are attributed to the interface and inner Fe atoms. Large V induced magnetic moments for the trilayer V/Fe/V were obtained by Clavero et. al. using a diverse assortment of experimental techniques [6]. These values, which range from -0.46 to -0.86 $\mu_B$ and are larger than the reported values[7] are attributed to interface mixing.

The earlier measurements on Fe/V superlattices using CEMS and vibrating-sample magnetometry showed that the Fe-V interface is paramagnetic [8]. From CEMS measurements, iron layers of thickness less than five monolayers were reported to be non-magnetic [9]. Kalska et al, using CEMS also reported differences in the degree of flatness between Fe on V and V on Fe interfaces [10].

Introduction of H into Fe/V superlattices strongly distort the local lattice structure as determined through extended x-ray absorption fine structure where variations up to around 7% of the local *c* parameter were reported [11-12]. Moreover, a hydrogen depletion zone of 1-2 monolayers in the V layer was reported [13] The magnetic properties of Fe/V superlattices can be reversibly tuned by hydrogenation [14]. For a two-monolayer Fe superlattice, it was found that the Fe atoms in the pristine sample possess a magnetic moment of 1.7 $\mu_B$, while for the hydrogenated samples the Fe atoms with neighboring V-H complexes possess moments in the range 2.1 – 2.2 $\mu_B$ [14]. The average magnetic moment per Fe atom was found to increase with increasing H content [15]. To our knowledge, there are no Mössbauer data on hydrogenated Fe/V superlattices. However, data on FeV alloys shows that addition of H transforms the paramagnetic FeV alloys to magnetic [16].

The location of H at interstitial sites in doped ultra-thin films of V was studied by Lebon *et al* [17] using the linear combination of pseudoatomic orbitals. Hydrogen was found to prefer the tetrahedral sites for substitutional transition-metal dopants located on the left of V in the periodic table, while octahedral sites are preferred for dopants on the right. Meded and Mirbt [18] studied the H loading in superlattices of X/V, (X = Cr, Mo, Mn, Fe) using the projector augmented waves. They found that H resides within the V layers out side the interface region with total energy trends following the accompanying volume expansion. The effect of H on the IEC and magnetic moment of a superlattice composed of three Fe layers and five V layers was studied by Ostanin et.al [19] using the full-potential linear-muffin-tin orbitals. The disappearance of the antiferromagnetic IEC for large H concentration was attributed to the decrease of the density of states at the Fermi level

We used the FP-LAPW method to calculate the hyperfine properties at the interface and inner Fe sites in $Fe_3V_9$ superlattice versus the H location and H content. It is found that sizeable changes in the hyperfine properties are obtained only when H is in the interface region.

## 2. Calculation method

Figure 1 shows the unit cell used to calculate the magnetic and hyperfine properties of the hydrogenated Fe/V superlattice $Fe_3V_9$. We have performed calculations with H atoms at both Oz octahedral and tetrahedral interstitial sites. The figure shows H atoms at the Oz octahedral sites. The H atoms are indexed by numbers from -1 to 5 with H at the interface Fe monolayer being indexed 0. Hydrogen with index 5 lies at the centre of the V layer, while those with indices 1 and -1 lie at the V interface monolayer and the Fe interior monolayer respectively.

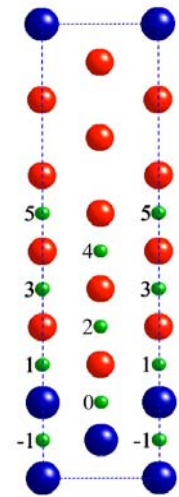

**Fig. 1**. The unit cell of the $Fe_3V_9$ superlattice. Dark (blue) large balls represent Fe and light (red) large balls represent V. The small (green) balls represent H at the (Oz) octahedral interstitial sites. The indices give the hydrogen locations within the superlattice.

The density functional L/APW+lo as implemented in WIEN2k package[20] is used to calculate the electronic and magnetic structures of FeV systems. In the L/APW+lo method the Kohn-Sham orbitals are expanded in terms of atomic orbitals inside the atomic muffin-tin (MT) sphere of radius $R_{MT}$ and in terms of plane waves in the interstitial regions. More detail on the calculation is found in Elzain and Al-Barwani [21].

## 3. Results and discussion

In table I we show the magnetic hyperfine fields ($B_{hf}$) at the interface and the central Fe monolayers together with the local Fe moments for different H locations. As seen from the table the values of $B_{hf}$ for H locations farther from the interface remain almost the same as that of the H free case $Fe_3V_9$. The magnitudes of $B_{hf}$ at both Fe sites for H at positions 1 and 0 at the interface are larger than that of $Fe_3V_9$.

Table I. The first and second rows give the contact hyperfine fields and the third and fourth row give the local magnetic moments at the interface Fe (0), the interior Fe (-1) for different H locations.

|         | $Fe_3V_9$ | -1    | 0     | 1     | 2     | 3     | 4     | 5     |
|---------|-----------|-------|-------|-------|-------|-------|-------|-------|
| Fe (0)  | -17.5     | -14.1 | -22.0 | -21.4 | -17.5 | -18.1 | -18.3 | -18.0 |
| Fe (-1) | -22.0     | -28.5 | -25.8 | -24.4 | -22.0 | -22.9 | -23.0 | -22.4 |
| Fe (0)  | 1.57      | 1.57  | 2.03  | 1.86  | 1.63  | 1.63  | 1.59  | 1.54  |
| Fe (-1) | 2.55      | 2.59  | 2.42  | 2.54  | 2.58  | 2.54  | 2.56  | 2.55  |

The corresponding local magnetic moments within the muffin-tin radii show almost similar trends. However, we find that the ratios of $B_{hf}$ to the local moments of Fe at the two sites are different. This makes doubtful the magnetic moments deduced from the experimental $B_{hf}$ or vise-versa using the same proportionality constant [22]. It is clear that the effect of H on the magnetic hyperfine fields is felt only when the H atom is located at the interface. On the other hand, Uzdin and Häggström used the floating point model that is based on the periodic Anderson model to study the interface properties through the distribution of the calculated magnetic moments. They concluded that comparison of the magnetic hyperfine fields distributions extracted from Mössbauer measurement and the calculated distributions of the magnetic moments support their proportionality [23].

The isomer shifts at the interface and central Fe atoms relative to α-Fe are shown in table II. These are calculated from the contact charge densities using a constant of proportionality of $-0.244\,a_0^3$ mm/s [24]. In general the values of the isomer shifts for H position far from the interface are small. Comparing the values of the isomer shifts to that of H free case ($Fe_3V_9$), sizeable changes are observed only when H is at the interface or inside the Fe layer. We found that the changes in the contact charge densities at the interface and the central sites follow the same trends exhibited by their change in volume as depicted in figure 2. This confirms our earlier conclusion that changes in the properties of Fe/V superlattices are attributed to volume effect [21].

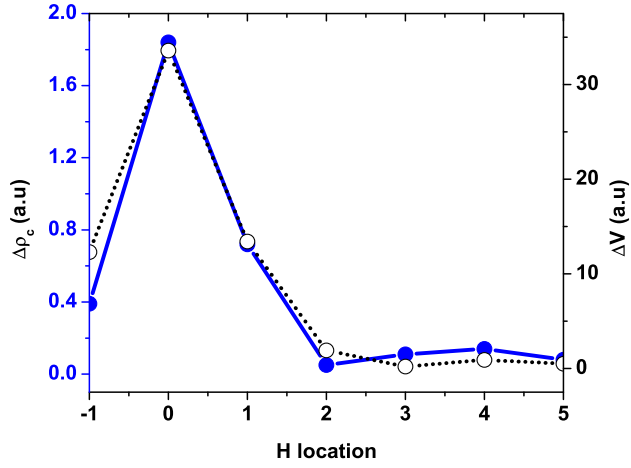

**Fig. 2** (a). Change in the contact charge density $\Delta\rho_c$ (solid curve) and the change in volume $\Delta V$ (dotted curve) at Fe interface atoms versus H location as indexed by monolayers starting with 0 at the Fe interface monolayer

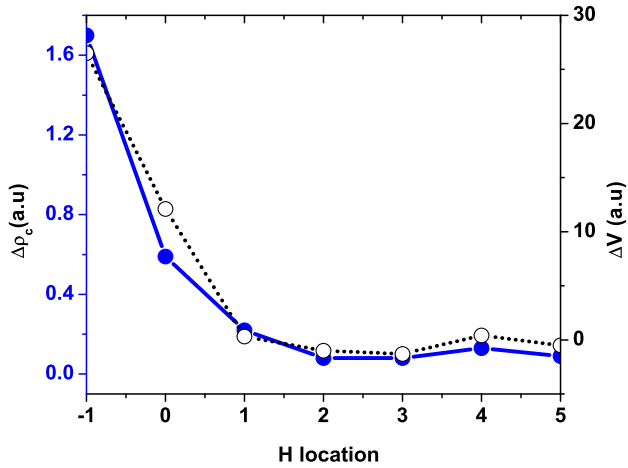

**Fig. 2** (b). Change in contact charge density $\Delta\rho_c$ (solid curve) and change in volume $\Delta V$ (dotted curve) at Fe central atoms versus H location as indexed by monolayers starting with 0 at the Fe interface monolayer

The quadrupole splittings shown also in table II in mm/s are in general large at the Fe interface atom and relatively small for the central Fe atom. However, significant changes from the H free case are observed for the interface atoms.

**Table II** The isomer (top) and quadrupole (bottom) splitting in mm/s at the interface Fe(0) and the central Fe(-1) atoms respectively for different H locations

| $Fe_3V_9$ | -1 | 0 | 1 | 2 | 3 | 4 | 5 |
|---|---|---|---|---|---|---|---|
| Fe (0) | -0.08 | 0.02 | 0.37 | 0.10 | -0.07 | -0.05 | -0.04 | -0.06 |
| Fe (-1) | 0.01 | 0.43 | 0.16 | 0.07 | 0.03 | 0.03 | 0.05 | 0.04 |
| Fe (0) | -0.40 | -0.40 | -1.24 | 0.09 | -0.33 | -0.60 | -0.32 | -0.41 |
| Fe (-1) | -0.05 | -0.79 | -0.09 | 0.09 | -0.12 | -0.05 | -0.07 | -0.06 |

From these results one concludes that the effect of H outside the interface region can not be detected by the Mössbauer measurement. However, this is true for a single H atom. We now consider the effect of introducing more than one H atom.

From the energetics of H occupation, the filling of Oz sites with H should start from the center of the V layer and proceeds towards the interface [18]. In table III we show the magnetic hyperfine field, the isomer shift and the quadrupole splitting at the interface and the central Fe atoms versus the H fraction (H/V) at the Oz site. We note that the isomer shift does not practically change with H contents, while the magnitude of the magnetic hyperfine field at central monolayer slightly increases. The magnitude of the quadrupole splitting decreases at first and then increases as H contents increases.

Table III The respective magnetic hyperfine field (in Tesla, top), the isomer shift (in mm/s, middle) and the quadrupole splitting (in mm/s, bottom) at the interface Fe(0) and the central Fe(-1) atoms versus the fraction of H at the Oz sites.

| H/V    | 0     | 1/9   | 2/9   | 3/9   | 4/9   | 5/9   |
|--------|-------|-------|-------|-------|-------|-------|
| Fe (0) | -17.5 | -18.0 | -18.6 | -18.9 | -18.7 | -18.5 |
| Fe (-1)| -22.1 | -22.4 | -22.9 | -22.9 | -23.1 | -23.6 |
| Fe (0) | -0.08 | -0.06 | -0.05 | -0.04 | -0.04 | -0.06 |
| Fe (-1)| 0.01  | 0.04  | 0.04  | 0.05  | 0.05  | 0.04  |
| Fe (0) | -0.40 | -0.41 | -0.36 | -0.34 | -0.52 | -0.59 |
| Fe (-1)| -0.05 | -0.06 | -0.06 | -0.10 | -0.05 | -0.05 |

As a result of these finding, when H is introduced in Fe/V superlattices there will be minor changes in the magnetic hyperfine fields and the quadrupole splitting at the Fe sites in the interface region if H is restricted to region outside the interface. This is despite the increase in the average magnetic moment with H contents observed experimentally[15] and confirmed theoretically[21]. However, the presence of H in the interface region will be reflected in large changes in all hyperfine parameters. Henceforth, Mössbauer measurements could be used to detect the absence or presence of H in the interface region.

In conclusion, the magnetic hyperfine fields, isomer shifts and quadrupole splittings at the interface and the central Fe sites in a $Fe_3V_9$ flat interface superlattice were calculated for different H locations and different H fillings. It is found that significant changes in the hyperfine parameters are detectable only when H resides in the interface region. This could be used to confirm the presence or absence of a depletion region in Fe/V superlattices.